# Enhancing Cardiovascular Disease Risk Prediction with Machine Learning Models

Farnoush Shishehbori[1], Zainab Awan[2]

**Affiliations**

1 William Harvey Research Institute, Barts and The London School of Medicine and Dentistry, Queen Mary University of London, London, United Kingdom.

2 William Harvey Research Institute, Clinical Pharmacology and Precision Medicine, Barts & the London Medical School, Queen Mary University of London, London, United Kingdom.

## Abstract

Cardiovascular disease remains a leading global cause of mortality, necessitating accurate risk prediction tools. Traditional methods, such as QRISK and the Framingham heart score, exhibit limitations in their ability to incorporate comprehensive patient data, potentially resulting in incomplete risk factor consideration. To address these shortcomings, this study conducts a meticulous review focusing on the application of machine learning models to enhance predictive accuracy.

Machine learning models, such as support vector machines, and Random Forest, as well as deep learning techniques like convolutional neural networks and recurrent neural networks, have emerged as promising alternatives. These models offer superior performance, accommodating a broader spectrum of variables and providing precise subgroup-specific predictions.

While machine learning integration holds promise for enhancing risk assessment, it presents challenges such as data requirements and computational constraints. Additionally, large language models have revolutionised healthcare applications, augmenting diagnostic precision and patient care.

This study examines the core aspects of cardiovascular disease event risk and presents a thorough review of traditional and machine learning models, alongside deep learning techniques, for improved accuracy. It offers a comprehensive survey of relevant datasets, critically compares ML models with conventional approaches, and synthesizes key findings, highlighting their implications for clinical practice.

Furthermore, the potential of machine learning and large language models in cardiovascular medicine is undeniable. However, rigorous validation and optimisation are imperative before widespread application in healthcare. This integration promises more accurate and personalised cardiovascular care.

## 1. Introduction

Cardiovascular diseases (CVDs) stand as prominent contributors to worldwide mortality. The World Health Organization (WHO) reports a staggering annual death toll of approximately 17.9 million due to heart-related issues and heart diseases. Notably, more than 80% of CVD-related deaths arise from heart attacks and strokes (World Health Organization, 2022). Therefore, detecting individuals at higher risk of CVD during its early stages and delivering suitable interventions can effectively prevent unexpected and early fatalities.

Cardiovascular disease, characterized by disorders affecting the heart and interconnected blood vessels, primarily arises from atherosclerosis—an inflammatory process involving plaque build-up with cholesterol, fat, and calcium, resulting in luminal narrowing and compromised blood flow. This condition impacts both the cardiovascular and cerebrovascular systems, contributing to ischemic stroke. Diagnostically, CVD is classified into four main areas (Olvera Lopez et al., 2023; Wilson, 2023). Coronary Artery Disease (CAD), the first category, is marked by insufficient blood supply to the myocardium due to epicardial coronary artery obstruction (Kannam et al., 2021). CAD includes manifestations such as myocardial infarction, angina pectoris, and coronary-related deaths, accounting for a significant portion of all CVD cases. Other categories involve Cerebrovascular Disease (including stroke and transient ischemic attack), Peripheral Artery Disease (primarily affecting the limbs and causing intermittent claudication), and Aortic Atherosclerosis, a chronic inflammatory disease linked to thoracic and abdominal aneurysms (Berbari & Mancia, 2015).

Cardiovascular event risks are influenced by well-researched factors such as smoking, diabetes, physical activity, dyslipidemia, obesity, family history, and low socioeconomic status. These risk factors have been extensively studied, as seen in research conducted by the British Heart Foundation (British Heart Foundation, 2019) and the American Society for Preventive Cardiology (ASPC) (Bays et al., 2022).

In the medical field, various multivariable prediction models, including the Framingham Cardiovascular Risk Score, European Systematic Coronary Risk Evaluation (SCORE), QRISK, and the American Heart Association/Atherosclerotic Cardiovascular Disease (AHA/ASCVD) algorithms, have been developed to assess cardiovascular risk. These models typically incorporate information on traditional CVD risk factors such as age, sex, smoking status, blood pressure, diabetes, and cholesterol levels (You et al., 2023).

However, traditional models have limitations that may impact their accuracy. These models often do not fully consider additional risk factors that have been identified in recent research (Gambardella and Santulli, 2016). Incorporating genetic risk information, for instance, has been shown to enhance CVD prevention strategies (Sun et al., 2021; Mamouei et al., 2023). The limitations inherent in traditional models stem from the variability among conventional risk scores, lack of comprehensive patient health records, demographic restrictions, lack of external validation, limited age suitability, and the assumption of linear relationships between risk factors while disregarding potential interactions between them. These limitations collectively undermine the accuracy and reliability of these models. (Sofogianni et al., 2022; Verweij et al., 2019; You et al., 2023).

To overcome these constraints and improve cardiovascular risk prediction, Machine Learning (ML) techniques have emerged as a valuable tool. ML techniques make use of a broader array of pre-existing medical conditions and can adapt to include new variables (Weng et al., 2017). Also, ML excels in integrating and interpreting diverse multimodal data sources, such as electrocardiogram (ECG) readings, genetics, and metabolic profiles, for comprehensive healthcare insights and disease detection (MacEachern & Forkert, 2021). A branch of ML known as deep learning (DL), demonstrates the potential to advance the accuracy of cardiovascular risk prediction (Mohd Faizal et al., 2021; Ranka et al., 2020).

Furthermore, Large Language Models (LLMs) are carving their niche in healthcare applications. In particular, Foundation Models (FMs) are acting as a vital bridge connecting vast electronic medical record (EMR) data to clinical practice, holding the potential to revolutionize the healthcare sector. The incorporation of LLMs into healthcare, however, necessitates rigorous clinical trials to ascertain their effectiveness and safety (Wornow et al., 2023; Thirunavukarasu et al., 2023). This integration represents a pivotal step toward the transformation of healthcare through the fusion of advanced AI and medical practice.

Within the dynamic healthcare landscape, the incorporation of ML and advanced DL techniques into conventional cardiovascular risk prediction models presents a path toward more accurate, personalized, and efficient patient care in cardiovascular medicine. This transition signifies a compelling paradigm shift, empowering healthcare practitioners with the tools to enhance decision-making processes and elevate the quality of cardiovascular care (Slomka et al., 2017). The present study delves into this transformative potential and seeks to contribute valuable insights into the future of cardiovascular healthcare.

Moreover, the primary aim of this study is to review and address several key research questions explicitly tied to the context provided earlier. These questions include evaluating the limitations of current traditional tools for predicting CVD risk, reviewing the use of ML models to improve CVD risk prediction, assessing the potential of DL models in this context, identifying relevant datasets for CVD risk prediction, conducting a comparative analysis to benchmark ML models against traditional methods, examining the integration of ML with conventional models, understanding the limitations inherent to ML models, and offering insights that can guide future research endeavours. This study ultimately seeks to harness the potential of ML to revolutionize the field of cardiovascular medicine, overcoming current limitations and considering a broader spectrum of risk factors.

## 2. Cardiovascular Events Risk and Its Importance in Clinical Practice

According to the WHO, major behavioural risk factors contributing to heart disease and stroke are an unhealthy diet, lack of physical activity, tobacco consumption, and excessive alcohol intake (World Health Organisation, 2022).

The British Heart Foundation identifies three main categories of risk factors for heart diseases: behavioural factors like smoking, alcohol consumption, and physical inactivity; health-related factors such as hypertension, high blood cholesterol, diabetes, and obesity; and demographic factors, including age, sex (with a higher risk in younger men than women), family history of heart disease, and ancestral or racial heritage (British Heart Foundation, 2019).

In the updated edition of 2022, the American Society for Preventive Cardiology (ASPC) (Bays et al., 2022) introduced a comprehensive list of ten key risk factors for cardiovascular diseases. These factors encompass various aspects, including unhealthy eating patterns, lack of physical activity, dyslipidemia, pre-diabetic/diabetic conditions, hypertension, obesity, special considerations for distinct demographic groups (such as advanced age, racial/ethnic disparities, and gender differences), thrombosis (with potential influence from smoking), impaired kidney function, and genetic factors, including familial hypercholesterolemia. This expanded framework underscores the multifaceted nature of cardiovascular risk assessment, encompassing lifestyle choices, health conditions, and individual characteristics.

### 3. Datasets Commonly Used for CVD Risk Prediction

The following datasets are often used for ML models to assess cardiovascular risks as shown in Figure 1:

- Cleveland dataset

- Medical Information Mart for Intensive Care (MIMIC-III) database

- UK Biobank

- University of California, Irvine (UCI) dataset

- National Health and Nutritional Examination Survey (NHANES)

- Clinical Practice Research Datalink (CPRD)

**Cleveland Heart Disease dataset:** This holds significant importance in ML and cardiovascular research. Comprising 303 individuals and originally featuring 76 attributes, it is noteworthy that only 13 of these attributes are consistently referenced in published studies. The remaining features describe the effect of the heart condition. The binary target variable indicates whether heart disease is present or absent in the individual (Abbasi et al., 2023).

**MIMIC-III:** It is a comprehensive and freely accessible resource for medical research, particularly in the field of intensive care. It contains data collected from patients admitted to critical care units at a single medical center in Boston, Massachusetts. It encompasses a diverse range of clinical data, including vital signs, laboratory results, medications, imaging reports, and demographic information (Johnson et al., 2016).

**UK Biobank:** It is an extensive biomedical repository situated in the UK, encompassing health and genetic data gathered from approximately 500,000 participants aged 40-69 years (UK Biobank, 2023)

**NHANES:** It is a program conducted by the Centers for Disease Control and Prevention (CDC) in the U.S. It aims to assess the health and nutritional status of a representative sample of the population. NHANES data is crucial for understanding health indicators, chronic diseases, and nutritional habits in the U.S., serving as a valuable resource for public health research (CDC, 2023).

**UCI dataset**: Founded in 1987 by David Aha and collaborators at UC Irvine, the UCI ML Repository serves as a crucial asset for the ML community. It includes databases, domain theories, and data generators employed in the empirical analysis of machine learning algorithms (BOCK, 2020).

**CPRD:** The CPRD is a valuable research resource in England, covering data from nearly 7% of the population. CPRD has played a major role in numerous studies, including validating predictive models like QRISK2 and exploring the potential of ML in healthcare. It serves as a versatile platform for both retrospective and prospective research in public health and clinical studies, offering valuable insights and supporting healthcare advancements (Li et al., 2020).

**Figure 1:** Datasets Frequently Used in Machine Learning Studies

## 4. Traditional Risk Prediction Models for CVD
### 4.1. Framingham

The Framingham Heart Study (FHS), initiated in 1949, aimed to explore the epidemiology of hypertensive or arteriosclerotic heart disease in the people of Framingham, Massachusetts (Dawber et al., 1951). The study tracked CVD incidence in individuals without prior heart disease symptoms. The Framingham Offspring Study started in 1971, and extended the investigation to the next generation, exploring genetic and familial aspects of heart disease (Feinleib et al., 1975). The Third Generation Cohort, initiated in 2002, expanded the understanding of CVD and lung health across three generations (Splansky et al., 2007).

Wilson and colleagues (Wilson et al., 1998) developed a gender-specific predictive model to evaluate the relationship between blood pressure, cholesterol categories, and coronary heart disease (CHD) risk. Building on this, a new sex-specific risk function tool was developed in 2008 (D'Agostino et al., 2008) that assessed the risk of a first cardiovascular event in addition to the 10-year probability of CHD occurrences.

Since 2009, the Framingham project has been at the forefront of cardiovascular research. It has seamlessly integrated advanced imaging technologies, such as cardiac MRI and multidetector CT scans. This ongoing research, enriched by genetic studies and expansive genome-wide association data, has identified significant associations between genetic variants and a range of traits, including blood pressure, BMI, vascular risk factors, and diverse cardiovascular diseases (Andersson et al., 2019).

In fact, in later models, the Framingham investigators expanded their focus to consider a broader spectrum of cardiovascular conditions beyond just coronary disease. This includes CVD mortality, general heart disease, stroke (which includes transient ischemic attack), intermittent claudication (leg pain), and CHF. This broader approach highlights the importance of addressing a wide range of CVDs rather than solely focusing on coronary disease. However, a limitation of the Framingham Risk Functions is their development based on data from a specific population, potentially leading to an overestimation of risk in new patients (D'Agostino Sr. et al., 2013).

### 4.2. Pooled Cohort Equations Calculator

The Pooled Cohort Equations Calculator (PCE), introduced by the ACC/AHA in 2013 (Stone et al., 2014), aims to estimate the high-risk group for atherosclerotic cardiovascular disease (ASCVD) and guide pharmacotherapy initiation. Despite its advantage in predicting risk for both White and Black populations, the PCE has faced challenges in accuracy and generalizability.

Utilizing basic factors such as gender, age, total cholesterol, HDL-C, systolic blood pressure, hypertension therapies, T2DM history, race, and smoking habits, the PCE estimates the 10-year risk of an individual experiencing their first severe cardiovascular event, including non-fatal heart attacks, death from CHD, and fatal or nonfatal strokes.

However, subsequent analyses revealed that the PCE significantly overestimated ASCVD risk in a contemporary, multi-ethnic population, leading to recommendations for recalibration (Rana et al., 2016; DeFilippis et al., 2015). A modified version introduced later aimed to improve accuracy, but evaluations suggested inadequate calibration and discrimination in real-world scenarios (Campos-Staffico et al., 2021).

While the PCE demonstrates effective risk discrimination in Hispanic individuals, its accuracy in predicting observed 10-year ASCVD events in multi-ethnic US cohorts, including the Dallas Heart Study (DHS) and the MESA, is less than optimal (Flores Rosario et al., 2021). The PCE may undervalue risk in certain racial and ethnic groups, potentially leading to insufficient pharmacotherapy in communities like Asian Americans, American Indians/Alaska Natives, and some Hispanics (DeFilippis et al., 2015).

Studies indicate that the PCE tends to overestimate cardiovascular risk, especially in the elderly (Nanna et al., 2019), and in overweight and obese individuals (Khera et al., 2020). The generalizability of the PCE among overweight individuals is questioned due to the lower proportion of individuals with obesity in the cohorts used for its derivation compared to present-day populations. Additionally, the PCE's overall performance is suboptimal in non-alcoholic fatty liver disease, especially in women and individuals with moderate-to-severe steatosis (Henson et al., 2022).

### 4.3. QRISK

The original QRISK model (Hippisley-Cox et al., 2007), designed to calculate the 10-year CVD risk, was released in 2007. The research encompassed a substantial UK primary care population, utilizing version 14 of the QResearch database[1]. This database, spanning 17 years and comprising the health records of 10 million patients from 529 general practices employs the EMIS computer system (a robust and validated electronic repository representative of primary care). In 2008, an updated version known as QRISK2 was released. QRISK2 consisted of extra risk factors such as ethnic origin, T2DM, rheumatoid arthritis, AF, and chronic renal disease. The age range for which QRISK2 is applicable has also been extended from 35-74 years to 25-84 years (Hippisley-Cox, et al., 2017).

The QRISK3 score was introduced in 2017 as an update to the QRISK2 algorithm. It was developed based on retrospective electronic health records (EHRs) from a large cohort of 2.67 million individuals (Hippisley-Cox, et al., 2017). QRISK3 introduces several new variables that are not present in the QRISK2 (You et al., 2023). These additions include an expanded definition of chronic kidney disease and seven additional factors that contribute to assessing the risk of CVD as shown in Figure 2 (NHS, 2021).

The QRISK2 score, validated in populations outside the UK, demonstrated accuracy (Collins et al., 2012). In 2021, QRISK3 was externally validated on CPRD (Livingstone et al., 2021). While the QRISK3 score is validated in the English population and performs well at the overall population level, it tends to overestimate 10-year CVD risk in elderly individuals and those with high multimorbidity. Validation studies for QRISK3 in non-English populations are pending, highlighting its effectiveness for the English population but raising considerations regarding its performance in the elderly and high multimorbidity categories.

---

[1] www.qresearch.org

**Figure 2:** Variables Incorporated in the QRISK Algorithms.

**Variables currently present in QRISK2-2017:**

- Age
- Ethnicity (nine categories)
- Townsend deprivation score
- Systolic blood pressure
- Body mass index
- TC/HDL-C ratio
- Smoking status can be categorized into different groups: non-smoker, former smoker, light smoker (1-9 cigarettes per day), moderate smoker (10-19 cigarettes per day), or heavy smoker (20 or more cigarettes per day).
- Family history of CHD in a first-degree relative before the age of 60
- Diabetes (type 1, type 2, or absence of diabetes)
- Treated hypertension (diagnosed hypertension requiring treatment with at least one antihypertensive medication)
- Rheumatoid arthritis (diagnosed rheumatoid arthritis, including related conditions like Felty's syndrome, Caplan's syndrome, adult-onset Still's disease, or unspecified inflammatory polyarthropathy)
- Atrial fibrillation (such as atrial flutter/fibrillation, and paroxysmal atrial fibrillation)
- Chronic kidney disease (stage 4 or 5) and major conditions like nephrotic syndrome, chronic glomerulonephritis, chronic pyelonephritis, requiring renal dialysis, or having undergone renal transplant (Hippisley-Cox, *et al.*, 2017).

**New or modified risk factors included in QRISK3:**

- Expanded characterization of chronic kidney disease: Includes a general practitioner-recorded diagnosis of chronic kidney disease stage 3, in addition to stages 4 and 5, as well as major chronic renal disease.
- Measurement of SBP variability: Takes into account the standard deviation of repeated blood pressure measurements.
- Migraine diagnosis encompasses a range of migraine types, including classic, atypical, basilar, hemiplegic and abdominal migraine, cluster headaches, and migraine with or without aura.
- Corticosteroid use: Consists of oral or parenteral prednisolone, betamethasone, cortisone, depomedrone, dexamethasone, deflazacort, efcortesol, hydrocortisone, methylprednisolone, or triamcinolone.
- SLE: Includes the diagnosis of SLE, disseminated lupus erythematosus, or Libman-Sacks disease.
- Use of second-generation "atypical" antipsychotics: Incorporates the use of medications such as amisulpride, aripiprazole, clozapine, lurasidone, olanzapine, paliperidone, quetiapine, risperidone, sertindole, or zotepine.
- Diagnosis of severe mental illness: Covers conditions like psychosis, schizophrenia, or bipolar affective disease.
- Diagnosis of HIV or AIDS
- Diagnosis or treatment of erectile dysfunction: Takes into account medications such as alprostadil, phosphodiesterase 5 inhibitors, papaverine, or phentolamine (Hippisley-Cox, *et al.*, 2017).

## 4.4. European Systematic Coronary Risk Evaluation (SCORE)

The SCORE is a tool for predicting the 10-year likelihood of cardiovascular death, developed using data from 12 European cohort studies (Conroy et al., 2003). It considers parameters such as age, sex, SBP, TC, and smoking status for individuals aged 40 to 65. A version for individuals aged 65–79, called SCORE Older Persons (SCORE O.P.) (Cooney et al., 2015), was introduced but exhibited insufficient discriminative ability, overestimating risk in those aged 65-69 and underestimating it in hypertensive patients aged 70-79 (Verweij et al., 2019).

The original SCORE model focused solely on fatal cardiovascular events and didn't consider non-fatal incidents, potentially underestimating overall CVD risk. This led to the development of SCORE2 in 2021, which predicts a 10-year risk for fatal and non-fatal CVD in individuals aged 40–69 without CVD or diabetes history (S.W Group and E.S.C.C.R Collaboration, 2021). SCORE2-OP was designed for those aged 70 and above, and SCORE2-Diabetes for individuals with type 2 diabetes (S.O.W Group and E.S.C.C.R Collaboration, 2021; S.D.W Group and E.S.C.C.R Collaboration, 2023).

The key strengths of SCORE2 lie in its foundation on extensive cohort studies conducted across diverse European countries, the availability of country-specific versions based on local data, and the heightened accuracy achieved through the integration of multiple data sources. This accuracy is further enhanced by considering contemporary occurrences specific to age and sex and distributions of risk factors, to make SCORE2 align with the latest information (S.W Group and E.S.C.C.R Collaboration, 2021).

SCORE2 was found more effective in recognising a larger percentage of rheumatoid arthritis patients as being at high or very high risk of CVD in comparison to the original SCORE. (Ferraz-Amaro et al., 2022). In predicting obstructive coronary artery disease, SCORE2 & SCORE2-OP showed independent predictive value for 4-year composite endpoints, outperforming Framingham. The study suggests enhancing the SCORE2 & SCORE2-OP prediction models by incorporating additional factors, such as the number of obstructive epicardial vessels and revascularization (Qiu et al., 2023).

A 'CKD Add-on' approach (Matsushita et al., 2022) was introduced to enhance the performance of the SCORE2 and SCORE2-OP models. This method involves integrating quantitative measures of CKD, specifically glomerular filtration rate and albuminuria, into the existing risk prediction models. The CKD Add-on exhibited superior performance compared to the original SCORE2 and SCORE2-OP models.

However, external validation in the Netherlands revealed an underestimation of CVD risk in low-risk countries like the Netherlands, particularly in subgroups with low socioeconomic status and Surinamese ethnicity (Kist et al., 2023). While SCORE2 represents a significant improvement, continuous research is essential for a comprehensive evaluation of its performance, particularly in diverse populations and subgroups. Ongoing efforts, such as the introduction of specialized versions like 'CKD Add-on' and the inclusion of additional risk factors, reflect the ongoing efforts to refine and enhance the predictive capabilities of this model.

### 4.5. Reynolds Risk score

The Reynolds Risk score (Ridker et al., 2007) customizes cardiovascular risk algorithms for women by incorporating additional variables, hsCRP, and parental family history of premature CHD. This augmentation, when integrated into the existing NCEP ATP III global risk score variables, enhances the accuracy of predicting CVD risk in women. The study followed participants for approximately 10 years, this predictive model incorporated risk factors including age, sex, SBP, HbA1c (if diabetic), smoking status, TC, HDL-C, as well as high-sensitivity C-reactive protein (hsCRP) levels and parental history of MI before 60 years old.

The Reynolds risk score showed superior performance compared to the Adult Treatment Panel III prediction scores in stratifying women initially classified as having an intermediate risk of CVD. The Reynolds data was derived from Women's Health Study (WHS), a nationwide cohort of US women, who were 45 years of age or older. The model successfully reclassified 40-50% of these women into either lower or higher-risk categories. The predicted outcomes from the Reynolds risk score aligned well with the actual events that occurred. In 2008, a male-specific Reynolds risk score (Ridker et al., 2008) was also created using data from a US cohort study.

However, it falls short in addressing unique risk factors for women, such as SLE, rheumatoid arthritis, preeclampsia, eclampsia, menopausal and hormonal changes, higher rates of depression and anxiety, and polycystic ovarian syndrome. (Isiadinso & Wenger, 2017).

While it demonstrates enhanced short-term predictive accuracy, the reliance on a specific cohort and modest overall gains suggests the need for continuous evaluation and comprehensive risk assessment to ensure broader applicability.

### 4.6. World Health Organization/International Society of Hypertension (WHO/ISH) Models

The WHO/ISH model (World Health Organization., 2007) consists of two sets of charts applicable in settings with or without blood cholesterol measurements, covering 14 WHO epidemiological sub-regions. These charts assess the 10-year risk for major cardiovascular events, incorporating factors such as age, sex, blood pressure, smoking, total cholesterol, and diabetes. A study in the Sri Lankans (Thulani et al., 2021) validated the charts, it was found suitable for use. However, it was observed that these risk charts exhibited better accuracy in estimating cardiovascular risk for males than for females. Furthermore, the risk charts demonstrated greater effectiveness in predicting risk among individuals classified as lower risk for cardiovascular events. However, another study in an Asian population (Selvarajah et al., 2014) showed that the WHO/ISH model performed poorly compared to the Framingham and SCORE models. The WHO updated its CVD risk prediction model in 2019 (WHO CVD Risk Chart Working Group, 2019) to better suit low-income and middle-income nations, with potential applications for diverse populations globally.

### 4.7. ASSIGN

The ASSIGN cardiovascular risk score, developed at Dundee University in 2006 (ASSIGN, 2014), estimates the 10-year risk of cardiovascular events in individuals without established CVD. It incorporates additional factors like social deprivation and family history, beyond those in the Framingham score. Based on data from the Scottish Heart Health Extended Cohort and is online-accessible. When compared to the Framingham and QRISK scores, the ASSIGN score has shown slightly better performance than the former (De La Iglesia et al., 2011). However, its use should be limited to the Scottish population, as it has not been validated outside of Scotland (Sofogianni et al., 2022).

### 4.8. PROCAM

The PROCAM risk score, established in Germany (Assmann et al., 2002), assesses the 10-year risk of acute CHD events using eight established risk factors. Initially designed for men aged 35 to 65, it was later updated to include both men and women for predicting coronary events and ischemic stroke (Assmann et al., 2007). However, a study revealed low sensitivity in detecting vascular age ≥70, particularly among women, emphasizing the need for accurate risk identification (Romanens et al., 2019). In the context of young HIV-infected males, PROCAM exhibited inferior performance compared to SCORE and DAD (Data Collection on Adverse Events of Anti-HIV Drugs) risk equations (Pirš et al., 2014).

The absence of validation beyond Germany and the limited scope of cardiovascular events covered by the PROCAM risk score indicate a need for external validation before its application. Also, it suggests potential limitations in accuracy, especially in certain demographics beyond the original study population.

### 4.9. CUORE Risk Score

The CUORE risk score, derived from the "Progetto CUORE" (HEART project), in Italy, predicts the 10-year likelihood of a first cardiovascular or cerebrovascular event aged 35 to 69 with no established CVD. The score incorporates eight recognized risk variables and is designed to estimate the probability of events like MI or stroke (Giampaoli et al., 2006; Doukaki et al., 2013).

The purpose of developing the CUORE risk score was to provide a more accurate estimation of the 10-year CVD risk specifically for the Italian people. However, it is important to note that the lack of external validation restricts the usage of the CUORE risk score to the Italian population.

## 5. Limitations of Current Cardiovascular Risk Prediction Tools
### 5.1. Overlooked Factors Associated with Cardiovascular Risks

Various multivariable prediction models have been developed and published to assess CVD risk, such as Framingham, SCORE, QRISK, and AHA/ASCVD algorithms. These models typically consider factors like age, sex, smoking status, blood pressure, diabetes, and cholesterol levels to estimate CVD risk. However, they might not fully account for other potential factors that could be associated with increased cardiovascular risk. Factors such as inflammation, Polygenic Risk Scores (PRSs) (Sun et al., 2021), vascular age (Terentes-Printzios et al., 2017), and Troponin I (Blankenberg et al., 2016) can play a crucial role in cardiovascular risk but are not consistently included in traditional models.

This limits their ability to accurately identify high-risk populations. Also, traditional methods might lack important information from patients' health records, potentially leading to overestimation or miscalculation (You et al., 2023). To improve risk estimation, additional risk factors have been proposed, such as vascular age (Terentes-Printzios et al., 2017), lipoprotein (A) (Bays et al., 2022) and genetic scores (Sun et al., 2021).

#### 5.1.1. Polygenic Risk Scores

Polygenic risk scores integration into CVD risk prediction models proves beneficial for identifying individuals at higher risk and enhancing risk discrimination and classification (Sun et al., 2021). Sun and colleagues' study, utilizing UK Biobank data, demonstrates significant improvements in the conventional risk prediction model when PRS information is added, preventing approximately 7% more CVD events compared to relying solely on conventional factors. Particularly efficient in individuals with intermediate risk.

Research by Khera et al. and Inouye et al. (2018) supports the effectiveness of PRSs in identifying individuals at significantly higher CVD risk. Integration of Inouye et al.'s PRSs with established risk factors showed a notable 3.7% improvement in CVD risk prediction discrimination. However, another study found minimal impact on risk discrimination with the addition of PRSs. Moreover, PRSs related to left ventricular cardiovascular magnetic resonance characteristics predict heart failure events independently of clinical risk factors (Aung et al., 2019), and a genetic score indicating albuminuria is significantly linked to an elevated risk of hypertension (Haas et al., 2018). Continuous updates are crucial given potential new genetic correlations (Sofogianni et al., 2022).

Limitations include a weaker evidence base for non-European ancestry (Lewis & Vassos, 2020) and challenges in interpreting PRS, which emphasise the need for nuanced assessments integrating genetic information with environmental factors.

### 5.2. Limitations of Conventional Risk Models in Personalized Risk Assessment

Conventional risk models fall short in providing personalized risk assessments because they primarily rely on data from larger populations, often overlooking unique individual variations in risk factors. Consequently, these models struggle to offer tailored risk evaluations to individual patients (Mohd Faizal et al., 2021).

### 5.3. Variability in Conventional Risk Scores

Conventional risk scores used for CVD assessment typically include common factors such as age, gender, smoking status, and hypertension. Nevertheless, variations exist in selecting parameters between different models, and, in some cases, these calculators differ in the level of detail and specificity of the questions asked as shown in Figure 3. For example, concerning family history, Reynolds mentions a parental family history of premature CHD, while PROCAM refers to the family history of MI. Smoking status, for instance, can be as straightforward as 'yes' or 'no' in one model, while another model may categorize it as former/light/moderate/heavy smoker. Additionally, diseases such as rheumatoid arthritis and chronic kidney disease are included in QRISK but not in others. Consequently, these differences result in significant variations in the calculated CVD risk between different calculators.

As an illustration, the Reynolds Risk Score demonstrates substantial differences when compared to the Framingham score across different risk categories. Table 6 provides detailed information, revealing that the Reynolds Risk Score identifies up to 70% more patients in the >20% risk category, 32% fewer patients in the 11–20% risk group, 5% more in the 5–10% group, and 28% higher numbers in the <5% risk category. Overall, the Reynolds Risk Score designates a high-risk group that is nearly twice as large (14%) compared to the Framingham score's high-risk group (8%) (Beswick et al., 2008).

Differences in eligibility criteria and outcome definitions across validation studies can impact the calibration of these models, potentially leading to overestimation or underestimation of cardiovascular risk in diverse populations.

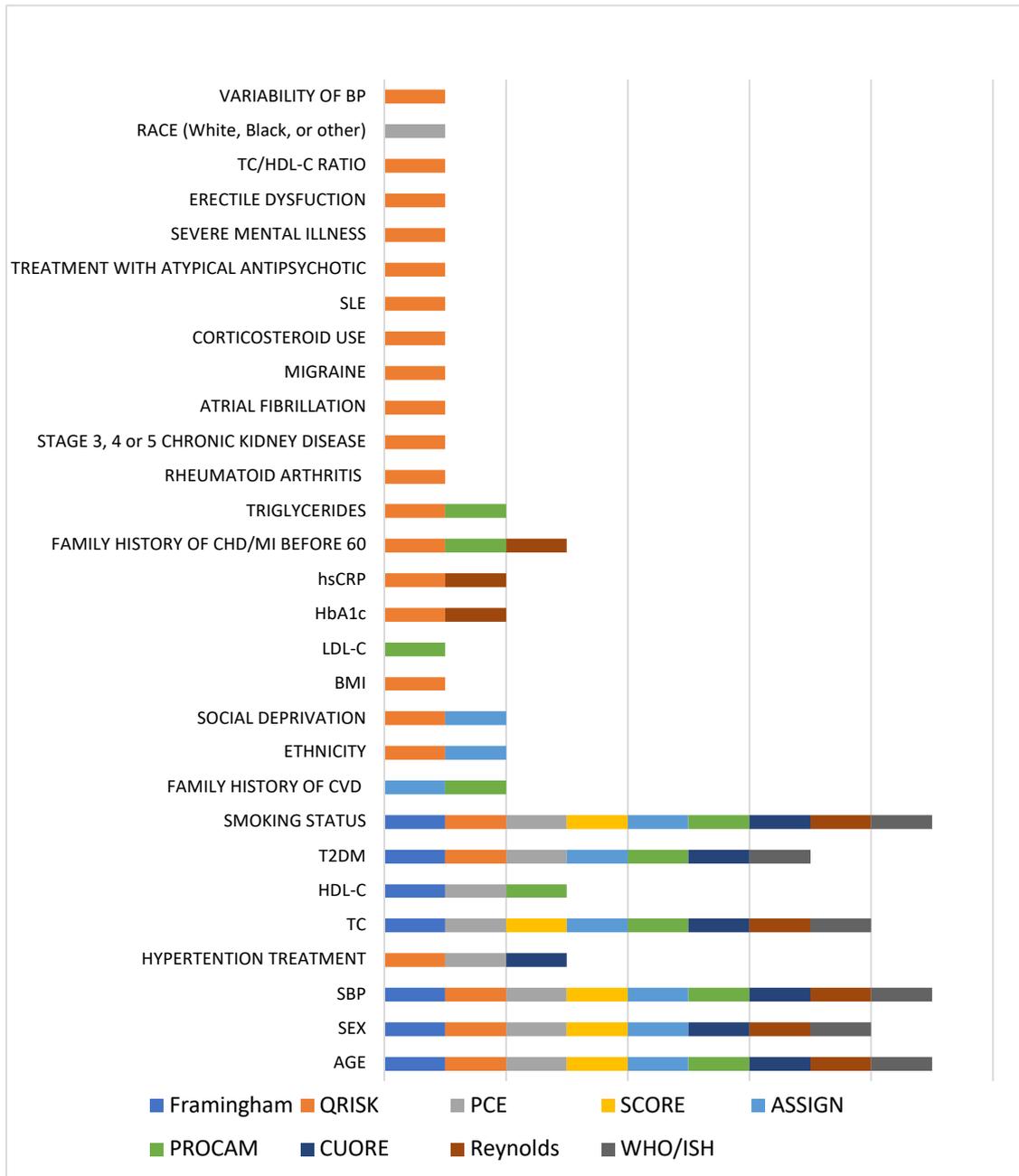

**Figure 3:** Comparison of Parameters Used to Estimate Cardiovascular Risks Among Different Models

The absence of ethnicity specification in risk predictions significantly affects the accuracy of individual risk assessments. For example, a 50-year-old person who lives and eats in a Western society with specific TC, HDL-C, and SBP values might be estimated to have a 4% risk according to the Framingham score. However, if he identifies his ethnicity as Indian American, the QRISK calculator would estimate the risk to be substantially higher, exceeding 20% (Abohelwa et al., 2023). Similarly, the Framingham risk assessment may substantially underestimate the risk of cardiovascular events for a 45-year-old non-smoking African American labour with moderate hypertension, no history of medical therapy, and a strong family history of MI, as compared to the QRISK criteria (Goff et al., 2014).

Although The Framingham Risk Functions generally worked well in predicting CHD events for white and African American populations (D'Agostino et al., 2001; Grundy et al., 2001), the functions have been found to overestimate CHD risk in Asian men from Honolulu and Hispanics from Puerto Rico, and required calibration adjustments for more accurate risk estimation. Therefore, adjustments to the risk functions are necessary to ensure accurate risk predictions for different populations (D'Agostino Sr. et al., 2013).

### 5.4. Linearity and Censoring

When developing multivariable prediction models, researchers traditionally used techniques like Cox or logistic regressions, which made assumptions about normal distribution, random censoring (when data is not complete), and linear correlations between predictors and outcomes. In practice, these assumptions could impact the accuracy and reliability of the predictive models. For example, certain risk factors may not follow a normal distribution, leading to skewed predictions. Moreover, random or uninformative censoring might result in incomplete data, potentially affecting the overall risk estimation. Additionally, many interactions are non-linear, and this linearity assumption could lead to inadequate representation of complex relationships between predictors and outcomes (You et al., 2023).

### 5.5. Lack of External Validation and Potential Overestimation or Underestimation of Risk in Diverse Populations

The lack of external validation in many cardiovascular risk scores raises concerns about their accuracy and generalizability when applied to populations beyond their original cohorts. To enhance accuracy in cardiovascular risk prediction, country-specific risk scores are preferable as they consider specific population characteristics and risk factors. However, the availability of validated country-specific risk scores may be limited for some regions (Sofogianni et al., 2022). A meta-analysis (Damen et al., 2019) examined the performance of three popular models—Framingham Wilson, Framingham ATP III, and PCE—in predicting a 10-year risk of CHD or CVD across diverse populations. Although overall performance differences were small, accuracy varied significantly depending on the population. All three models tended to overestimate risk if not recalibrated to the local context, with European populations experiencing more pronounced overestimation compared to the USA. Updating the models with local data improved performance, revealing better discriminative ability. The c-statistic displayed considerable heterogeneity across studies, emphasizing the need for external validation and local recalibration to ensure accurate and relevant cardiovascular risk prediction across diverse populations.

## 6. ML Models for CVD Risk Prediction

Machine learning, a subfield of artificial intelligence (AI), has gained popularity in cardiovascular medicine. ML is frequently recommended for forecasting heart disease due to its capability to derive information efficiently and precisely from extensive datasets, simplifying the prediction process. ML Models take input data and utilize mathematical optimisation and statistical analysis to anticipate outcomes. In fact, ML forms the fundamental basis that aids in managing vast data volumes, offering high-speed processing, and delivering early-stage predictions. ML can be classified into three distinct types of supervised, unsupervised, and reinforcement (Krittanawong et al., 2020; Azmi et al., 2022).

### 6.1. Supervised ML

Supervised ML holds a prominent position within the ML domain, finding extensive utility in cardiovascular medicine. This method relies on labelled data pairs, encompassing inputs and their corresponding outputs, to train models effectively. A variety of approaches is harnessed to achieve this proficiency, including the utilization of methods such as the naive Bayes theorem, support vector machines (SVM), tree-based strategies like random forests, and the power of neural networks (Kilic, 2020; Ranka et al., 2020). Among ML models, SVM and Random Forest, as well as GBM, have garnered significant interest among researchers. These applications are discussed in detail in Figure 4.

### 6.2. Unsupervised ML

Unsupervised ML is a subset of ML that identifies patterns in unlabeled datasets, aiming to reveal inherent relationships and patterns within the data (Ranka et al., 2020). This process involves automatically categorizing data into distinct subgroups or clusters based on similarities or patterns, unveiling hidden structures that may not be immediately apparent. This approach utilizes techniques such as clustering, which groups similar data points without predefined labels to identify natural clusters, and Principal Component Analysis (PCA), a dimensionality reduction method transforming high-dimensional data into a lower-dimensional space while retaining key variance. Notably, unsupervised ML holds great promise in applications like cardiovascular disease classification (Mathur et al., 2020; Zheng et al., 2020).

### 6.3. Reinforcement learning

Reinforcement learning, a combination of supervised and unsupervised learning, utilizes an iterative trial-and-error process to enhance algorithm accuracy in healthcare, particularly in cardiovascular care settings (Krittanawong et al., 2017). While not as widespread as other learning methods, reinforcement learning shows significant potential by providing continuous treatment recommendations and interventions. This includes personalized adjustments to medication dosages, lifestyle suggestions, and strategic interventions based on dynamic patient responses and evolving risk factors.

In essence, reinforcement learning trains a computational agent to optimize cumulative rewards over a sequence of time steps. The agent observes its surroundings and evaluates the outcomes of its actions in a specific situation. (Moazemi et al., 2023). Figure 4 illustrates various applications of diverse machine learning-based approaches in cardiovascular medicine.

### 6.4 Variability in ML Model Selection for Cardiovascular Disease Prediction

It is unpredictable to determine a single preferred ML method for CVD prediction. It is common practice for researchers to evaluate and compare different ML models to determine which one performs best for a particular CVD prediction task. The choice of the most appropriate model may vary from one study to another based on the context and the dataset being used. Therefore, there is no one-size-fits-all answer, and the selection of the ML method should be based on empirical evidence and suitability for the given problem (Bhadri et al., 2022).

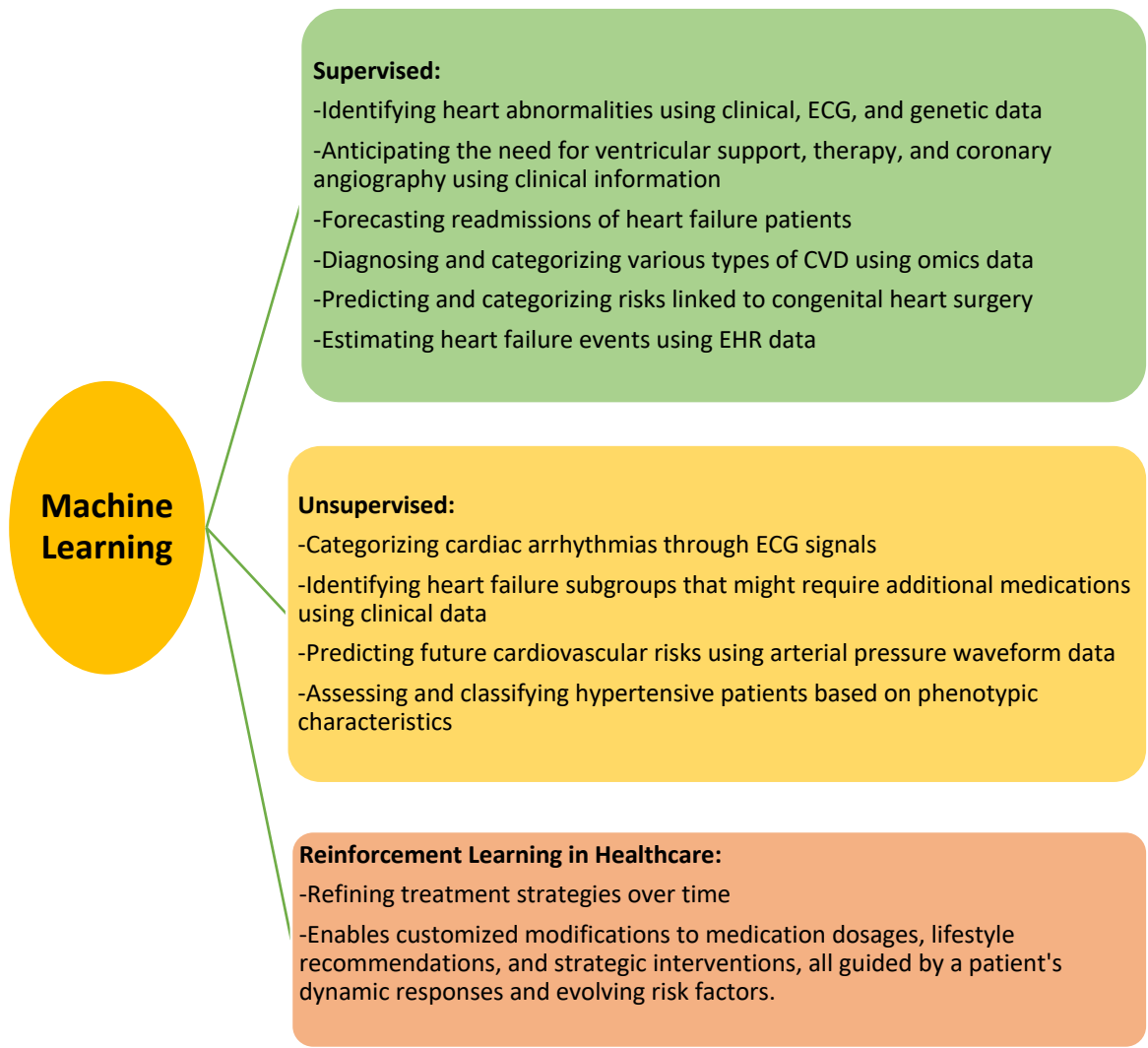

**Figure 4:** Applications of Different ML-Based Approaches in Cardiovascular Medicine (Mathur et al., 2020; Moazemi et al., 2023).

7. **Deep Learning**

Deep Learning (DL), a rapidly advancing subfield within ML, employs multi-layered neural network architectures (Mohd Faizal et al., 2021) characterized by their ability to process extensive datasets, uncover intricate connections, and create predictive models. DL's widespread exploration and acceptance in cardiovascular healthcare are crucial for its extensive application (Ranka et al., 2020). It distinguishes itself from traditional ML by utilizing more sophisticated computational models with multiple layers.

Autoencoders, a specific DL technique, play a vital role in simplifying complex data and reducing dimensionality without the need for labelled information, making it more manageable. It has the adaptability and intelligence to find patterns in data (Hinton & Zemel, 1994; Paixão et al., 2022). DL-based neural network algorithms, such as CNN and RNN, play vital roles in tasks such as arrhythmia detection(Xiong et al., 2018; Hannun et al., 2019), atrial fibrillation (AF) classification (Mahajan et al., 2020), CHD diagnosis (Acharya et al., 2018), ECG interpretation (Smith et al., 2019), coronary artery disease prediction (Betancur et al., 2019), cardiac Magnetic Resonance Imaging (MRI) analysis (Avendi et al., 2016; Tan et al., 2017), and quantifying atherosclerotic plaque (Lin et al., 2022). This advancement contributes to improved diagnostic accuracy, reduced interpretative errors, and enhanced risk assessment in cardiovascular healthcare.

Xiong et al. (2018) developed a DL-based model called RhythmNet, which combines CNN and RNN approaches to classify and diagnose AF using ECG data (Xiong et al., 2018).

In another investigation (Tison et al., 2018), deep neural networks (DNNs) were deployed to detect AF using data obtained from smartwatches. The research employed smartwatches to gather heart rate and step count data for algorithm development. Notably, the DNN underwent heuristic pretraining, allowing it to approximate representations of the R-R interval (the time between heartbeats) without the need for manual labelling. The trained DNN, employing this innovative approach, showcased a remarkable C statistic of 0.97 in detecting AF, outperforming the reference standard of the 12-lead ECG.

Furthermore, in cardiology research, the utilization of DL models in conjunction with fundus photography as an imaging modality has gained prominence (Chang et al., 2020; Son et al., 2020). For instance, Poplin and colleagues (Poplin et al., 2018) harnessed retinal fundus images to predict various cardiovascular risk factors such as age, gender, and smoking status, in addition to predicting major adverse cardiac outcomes. Their deep learning model, trained on fundus photography data, obtained an AUC of 0.70, and this performance was validated in two distinct cohorts.

RNNs, designed for sequential data handling, exhibit potential in identifying newly occurring cases of heart failure (HF) with notable AUCs of 0.777 and 0.883 for 12-month and 18-month observation windows, respectively (Choi et al., 2016).

The application of DL in cardiovascular health extends to the diagnosis of ECG-related conditions. Deep neural networks, particularly DNNs, excel in predicting one-year all-cause mortality using ECG data, surpassing traditional clinical risk scores and demonstrating predictive precision beyond one year (Han et al., 2023; Raghunath et al., 2020).

RNN models leveraging temporal relations enhance the detection of HF using Electronic Health Records (EHR) data, showing superior performance compared to conventional methods (Choi et al., 2017).

### 8. Strengths of ML in Comparison to Conventional Approaches

Conventional methods for CVD risk prediction, including widely used tools like ACC/AHA, QRISK2, Framingham, and Reynolds, have struggled to accurately identify individuals who could benefit from preventive treatment. This challenge leaves a significant number of people at risk of CVD undetected by traditional tools, while those with low risk may receive unnecessary preventive measures. Remarkably, almost half of heart attacks and strokes occur in individuals not expected to be prone to CVD (Weng et al., 2017).

Traditional CVD risk models assume linear relationships between individual risk factors and CVD outcomes, leading to the oversimplification of complex interactions among numerous factors with non-linear relationships. ML shows promise in addressing this challenge by capturing intricate interactions among various risk factors. (Obermeyer & Emanuel, 2016). Moreover, it can unveil new risk factors, allowing for more precise predictions tailored to individual sub-populations (Alaa et al., 2019).

One of ML's strengths lies in its proficiency in utilizing multimodal data, which has become increasingly available with advancements in medicine, genetics, metabolic studies, and imaging technologies (MacEachern & Forkert, 2021). ML is being applied to analyse extensive data from various sources, including physiological sensors, genomic sequencing, imaging tests such as nuclear stress tests, computed tomography, echocardiography, and electronic medical records (Krittanawong et al., 2017). ML techniques can process multimodal data to identify biomarkers associated with specific disease subcategories, enhancing early disease detection, predicting responses to medications, and offering insights into patient prognosis (MacEachern & Forkert, 2021).

## 9. Limitations and Pitfalls of ML:

Despite the potential benefits of ML in CVD risk prediction, there are significant limitations associated with its use. Some of the limitations and pitfalls of ML include:

### 13.1. Imbalanced Data

Imbalanced datasets, where one class dominates, can lead to biased model performance. Techniques like undersampling and oversampling, particularly using SMOTE, are employed to address this issue. Undersampling reduces the majority class while oversampling increases the minority class. SMOTE, a widely used method, enhances model performance metrics like AUC-ROC and accuracy (Pan et al., 2020). Recent advancements highlight Generative Adversarial Networks (GANs) as effective tools for managing imbalanced data by generating synthetic data resembling genuine data (Ahsan & Siddique, 2022).

### 13.2. Censoring

Censoring in time-to-event analysis refers to incomplete or partially missing data points, typically occurring when information on the time to the outcome event is not available for all study participants due to factors such as loss to follow-up or death (Prinja et al., 2010).

ML models, when applied without considering censoring, can introduce significant bias in risk predictions, especially in datasets with substantial censoring. Failure to account for censoring may lead to the incorrect assumption that censored participants are event-free, resulting in a notable underestimation of the actual CVD risk. An assessment of censoring's impact on risk prediction involved 3.6 million patients in England over 20 years (Li et al., 2020). The study revealed that models neglecting censoring, including common ML models, produced biased risk predictions, particularly for higher-risk patients. For example, in patients with a 9.5-10.5% risk according to QRISK3, the range of expected risk differences between QRISK3 and a neural network was found to be between -23.2% and 0.1% (Li et al., 2020).

### 13.3. Challenges in Variable Selection for Risk Prediction Models

An unanswered question in risk prediction revolves around determining the appropriate types and number of variables to include in models, as there is currently a lack of consensus and guidelines for variable selection. The inclusion of more predictors in a model can potentially raise the C statistic due to increased predictor variation. However, incorporating non-causal predictors might lower accuracy as this adds noise, increasing the risk of overfitting, along with associated challenges related to data quality (Li et al., 2020).

### 13.4. Overfitting and Underfitting:

Machine learning models have the potential to learn complex patterns in data, but it is crucial to avoid overfitting the training data, which can happen when the model is trained on a dataset that is too small or too similar to the test dataset, causing the model to learn too much about the particular training set and fail to generalize to new data. The risk of overfitting increases with the complexity of the model, especially in DL networks with lots of parameters. Conversely, models can also underfit the training data (failing to capture the data's complexity), resulting in reduced accuracy in both the training and testing datasets (MacEachern & Forkert, 2021).

### 13.5. Missing Data:

Missing data could include everything from variables with missing values to samples with unreported information. These can cause inaccuracy in the model. Of course, some ML techniques, such as Random Forest and K-Nearest Neighbour, are capable of handling missing values directly (Picard et al., 2021).

### 13.6. Interpretability:

Machine Learning models, especially those applied to multi-omics data, can be intricate and challenging to interpret due to the complex relationships between different omics layers. The latest DL methods, in particular, are often considered black-box models, making interpretation difficult. Neural networks used in these algorithms extract features in a way that is hard to trace back and understand, leading to a lack of clarity in their output (Wen et al., 2023). The interpretability challenge poses difficulties in identifying and correcting potential errors within the model (Murdoch et al., 2019).

Healthcare professionals often prefer ante-hoc interpretable models that are naturally understandable and intuitive (Xie et al., 2022) for instance, utilize a Directed Acyclic Graph (DAG) to visually represent the factorization of the joint probability distribution of the data.

This graphical representation explicitly shows the dependencies between input variables and the predicted outcome, making Bayesian models inherently interpretable (Lisboa et al., 2023).

### 13.7. Legal and Ethical Issues

The interpretability of ML models can raise ethical and legal issues, especially as the complexity of input data and models increases. Healthcare professionals may not fully understand how a model reaches a specific conclusion about a patient's care. While physicians can be held legally liable for their actions, it is not apparent who should be held accountable when an ML model makes a poor choice (MacEachern & Forkert, 2021).

## 10. Discussion

### 10.1. Comparison of the Performance of ML Models Against Traditional Methods

In a prospective study (Alaa et al., 2019) using the UK Biobank dataset, AutoPrognosis, an autonomous algorithmic tool. This approach is used to ensure that the most suitable ML model is utilized for the specific dataset and task, without requiring researchers to make pre-determined assumptions about which model would work best. AutoPrognosis achieved a significantly improved AUC-ROC of 0.774, compared to the Framingham score (0.724), the Cox PH model with conventional risk factors (0.734), and the Cox PH model using all UK Biobank variables (0.758). It identified new predictors, including non-laboratory variables, showcasing its ability to handle large-dimensional datasets and identify complex interactions for more precise predictions tailored to individual sub-populations, including those with diabetes. One significant marker for elevated CVD risk among individuals with diabetes was found to be urinary microalbumin.

In another prospective cohort study (Weng et al., 2017), involving 378,256 patients, ML algorithms (logistic regression, random forest, gradient boosting machines, and neural networks) demonstrated substantial improvements in predictive accuracy (ranging from 1.7% to 3.6%) compared to the ACC/AHA risk model (AUC: 0.728) for CVD risk prediction. ML models identified new risk factors not included in traditional tools, such as COPD, severe mental illness, and the prescription of oral corticosteroids. Neural Networks[2] showed a substantial 3.6% improvement in predictive accuracy, highlighting the superior proficiency of ML methods in identifying individuals prone to CVD development compared to traditional methods.

---

[2] The selected neural network model in the study is a single hidden layer neural network with the hyperparameters: size = 3 and decay = 0.09.

## 10.2. Improvement in Clinical Decision-Making with Integrated ML Models and Traditional Tools

The integration of an ML model with coronary CT angiography (cCTA)-derived plaque measures and clinical data demonstrates remarkable predictive capabilities for major adverse cardiac events (MACE). In a recent study (Tesche et al., 2020), the ML model, employing decision trees and the RUSBoost classification algorithm, outperforms traditional coronary CT risk scores and clinical parameters in predicting MACE over a median follow-up period of 5.4 years. The ML model's accuracy, with an AUC of 0.96, significantly surpasses that of conventional risk scores and adverse plaque measures, offering meaningful clinical implications for more precise risk management and tailored treatment strategies for individuals at risk of MACE.

In a study aimed at improving predictions related to atherosclerotic cardiovascular disease (ASCVD) and coronary heart disease (CHD) fatalities, Nakanishi et al. (2021) explored the impact of ML integration in cardiovascular medicine. The research focused on assessing whether ML models, trained on non-contrast CT and clinical data, could enhance predictive accuracy compared to traditional methods such as the ASCVD risk score and coronary artery calcium (CAC) Agatston scoring. Utilizing an ensemble boosting approach (LogitBoost) with 77 variables from clinical and CT data, the comprehensive ML model, referred to as ML all, outperformed other models in predicting CVD and CHD deaths, achieving AUCs of 0.845 and 0.860, respectively. This highlights the superior predictive accuracy of the ML approach and suggests its potential as a routine tool for clinical risk evaluation in cardiovascular medicine.

## 10.3. Transforming Cardiovascular Care: The Impact of ML on Clinical Decision-Making and Patient Outcomes

The integration of ML into clinical decision-making is introducing a new era of precision and effectiveness. Traditionally, cardiovascular care relied on subjective visual interpretations and conventional risk assessment tools. However, ML is introducing a data-driven, evidence-based approach that reduces subjectivity and improves precision in clinical decision-making. ML-driven automation and quantitative analysis minimize interpretation variability in imaging results and risk assessments. This objectivity is pivotal for clinicians, enabling them to make more informed choices in patient care.

ML's capacity for precise quantitative analysis fundamentally transforms cardiovascular risk assessment by rapidly and accurately processing extensive imaging data, leading to highly reliable measurements and risk predictions. This precision enhances the credibility and accuracy of cardiovascular risk assessment models, providing clinicians with invaluable tools for patient care (Slomka et al., 2017).

Also, ML demonstrated its proficiency to employ a broader range of variables and unveil new risk factors, resulting in more precise predictions tailored to individual sub-populations such as individuals with diabetes (Alaa et al., 2019). Moreover, the predictive abilities of DNNs go even beyond conventional boundaries. To illustrate, they perform exceptionally well when applied to ECGs that medical professionals had previously classified as within normal parameters, capturing intricate interactions, and unveiling patterns that traditional methods had overlooked (Raghunath et al., 2020).

Integration of ML with clinical data offers a thorough understanding of each patient's condition. Combining information from clinical records and imaging results enables a patient-specific approach to cardiovascular medicine. It supports not only risk assessment but also the development of tailored treatment plans, aligning care with each patient's unique needs.

Moreover, ML streamlines image processing and interpretation, significantly improving efficiency in clinical practice. While it does not replace healthcare professionals, it enhances the analysis process, reducing the time and effort required for interpretation. This newfound efficiency allows clinicians to focus on higher-level decision-making and patient interaction, optimizing their expertise and time resources (Slomka et al., 2017).

### 10.4. Evaluating Consistency of Predictive Models in Clinical Decision-Making

Before employing models for clinical decision-making, it is crucial to assess their consistency, both within and between different models. This evaluation should be integrated into the TRIPOD (Transparent Reporting of a multivariable prediction model for Individual Prognosis or Diagnosis) guidelines. These guidelines are designed for creating and validating risk prediction models but primarily emphasise how well these models perform for the entire population. Unfortunately, the consistency of individual risk predictions generated by multiple models, even if they perform similarly on a population level, is not addressed by these guidelines. In other words, before using these models to make medical decisions, we must ensure they not only work well on average but also provide consistent and reliable predictions for individual patients (Li et al., 2020).

## 10.5. Large Language Models

Large Language Models (LLMs) are foundational AI systems built upon transformers. They utilize a self-attention mechanism for comprehending input relationships. They undergo a two-step training process: pretraining involves self-supervised learning on unannotated data, without the need for manual labelling. This confers a notable advantage over traditional deep learning models that rely on full supervision, as it eliminates the need for extensive manual annotation. Subsequently, in the fine-tuning stage, LLMs are trained on small, task-specific annotated datasets, utilising the knowledge acquired in the pretraining phase to execute specific tasks as intended by the end user. Consequently, LLMs demonstrate high accuracy across diverse tasks with minimal human-provided labels (Shen et al., 2023; Singhal et al., 2023).

In terms of structure, LLMs encode raw textual input into tokens, representing words or characters, and transform them into high-dimensional vectors known as embeddings. These embeddings, acquired through exposure to extensive training data, undergo a sequence of transformer layers employing self-attention mechanisms to unveil intricate word relationships. The final step involves a linear layer that decodes insights and generates predictions based on collective knowledge as shown in Figure 5 (Reddy, 2023).

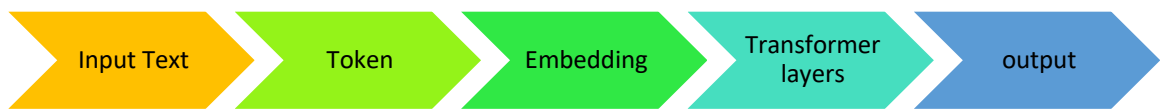

**Figure 5**: Structure of LLMs - Adapted from (Reddy, 2023)

### 10.5.1. Foundation Models in Healthcare:

In tandem with the advancements in the field of ML, LLMs are carving their niche in healthcare applications. In particular, Foundation Models (FMs) are acting as a vital bridge connecting vast EMR data to clinical practice, holding the potential to revolutionize the healthcare sector.

There are two main categories of Clinical FMs: Clinical Language Models (CLaMs) and Foundation Models for Electronic Medical Records (FEMRs).

CLaMs are a subtype of LLMs designed for clinical and biomedical text. They are trained to process and generate clinical information, making them suitable for tasks like extracting drug names from medical notes, answering patient questions, or summarizing medical dialogues. CLaMs that learn from clinical text primarily use a single database called MIMIC-III (which encompasses about 2 million ICU notes generated between 2001 and 2012). On the other hand, CLaMs trained in biomedical text typically rely on PubMed abstracts or complete research articles.

FEMRs focus on patient representations (a compact, multi-dimensional vector that compresses extensive patient data) rather than clinical text. They process a patient's entire medical history and produce machine-understandable patient embeddings. These representations can be used as input for various downstream models, enhancing accuracy in tasks like predicting readmission rates or hospital stay lengths.

### 10.5.2. LLM Applications in Disease Risk Prediction Models

However, it is crucial to note that LLMs are far from perfect. They have limitations and can make mistakes, particularly regarding specialised medical knowledge and individual patient circumstances. While there have been suggestions that AI could replace doctors, the reality is more nuanced. LLMs, including the latest GPT-4, have their strengths but also significant weaknesses. They may provide valuable assistance but cannot autonomously make critical medical decisions or communicate with patients.

LLMs in medicine face challenges related to accuracy, lack of up-to-date information, transparency, and ethical concerns (Thirunavukarasu et al., 2023).

A more ambitious application of LLMs is indeed in clinical cardiology decision support, but achieving this goal is a distant prospect. In the short term, LLM applications are centred around text-related tasks like generating automated documents and patient summaries (Sarraju, et al., 2023). Future research should include rigorous clinical trials to assess the real-world acceptance, effectiveness, and practicality of LLM medical applications. These trials are essential to ensure that LLM interventions meet medical standards and lead to improved patient outcomes (Thirunavukarasu et al., 2023).

## 11. Conclusion

In conclusion, traditional models for cardiovascular risk prediction exhibit various limitations, such as overlooking important risk factors, variability among conventional risk scores, a lack of comprehensive patient health records, and issues related to ethnicity, external validation, and assumptions of linear relationships. These limitations are evident in the existing risk prediction landscape.

Advancements in cardiovascular risk prediction show great promise, with the incorporation of PRSs holding the potential to enhance CVD risk prediction and prevention, offering notable improvements in risk discrimination and reclassification.

Furthermore, ML models have demonstrated their superiority over traditional methods by providing enhanced predictive accuracy, thanks to their ability to incorporate a wider range of variables, capture complex risk factor interactions, accommodate non-linear relationships, handle multimodal data, and offer improved subgroup-specific predictions. Yet, it is essential to recognize that choosing the most suitable ML method may vary based on the specific context and dataset used. Despite their benefits, ML methods face limitations, including issues like censoring, interpretability, imbalanced data, overfitting, underfitting, and ethical concerns, as well as challenges in variable selection.

On the other hand, DL techniques, including deep neural networks and RNN, demonstrate superior performance in CVD risk prediction compared to traditional ML classifiers. These approaches are particularly effective in tasks, such as image analysis, where CNNs shine, offering significant enhancements in areas like CT images, coronary image segmentation, and ECG interpretation. Nonetheless, the adoption of DL in CVD research faces challenges, such as data requirements, expertise gaps, and computational constraints.

Moreover, upon validation of ML algorithms' superiority over traditional risk models through head-to-head comparative studies, the most efficient algorithms can be seamlessly integrated into EHR for practical deployment in clinical environments.

The integration of ML into clinical decision-making is revolutionizing cardiovascular care by introducing precision and objectivity. ML models can offer precise risk assessment and tailored treatment strategies for individuals at risk of major cardiac events.

Additionally, the use of LLMs adds another layer of sophistication to prediction tasks, outperforming traditional models with heightened sensitivity and specificity. These LLMs excel in natural language understanding and generation tasks, offering cost-effective, multimodal data handling, and demanding less labelled training data. They serve as foundational models that reshape the landscape of medical research and clinical practice, with significant potential for enhancing diagnosis, prognosis, and patient care.

Despite these advancements, it is crucial to ensure the consistent and accurate performance of these models on an individual patient level and to conduct rigorous clinical trials before widespread implementation. The integration of ML models, PRSs, and LLMs into clinical practice offers substantial benefits but must be carefully validated and refined for optimal utility in healthcare.